# The Thermal Regulation of Gravitational Instabilities in Disks Around Young Stars
## (Conclusions Chapter)*

**Annie C. Mejía**

*Indiana University, acmejia@astro.indiana.edu*

The work presented in this dissertation is aimed at understanding how cooling and heating processes affect the development of gravitationally unstable protoplanetary disks. The three dimensional hydrodynamic simulations presented in Chapters 3 and 4 model a relatively massive (0.07 $M_\odot$) self-gravitating disk, which surrounds a typical pre-main-sequence solar-type (0.5 $M_\odot$) star. The disk extends 40 AU initially, but eventually expands as far as 85 AU from the central star after the gravitational instabilities develop. This model probably corresponds to a few $\times 10^5$ year old disk.

## 1: Motivations and Goals

Studies by various groups on the hydrodynamics of circumstellar disk have shown that gravitational instabilities are sensitive to the assumed equation of state of the gas and that their strength, longevity, and tendency to fragment are entirely regulated by thermal processes (Tomley et al. 1991, 1994; Pickett et al. 1998, 2000a,b; Gammie 2001). This in turn implies that the amount of restructuring that occurs in a disk as a result of it becoming gravitationally unstable and the efficiency of GIs as a planet formation mechanism are also subject to energy flow rates.

In order to move toward more realistic treatments of cooling and heating processes in a gaseous disk, the hydrodynamics code uses an ideal gas law replacing simplistic equations of state and adding irreversible heating by shocks (Pickett et al. 1998, 2000b). The further addition of algorithms to simulate irradiation, radiative cooling and diffusion make the code considerably more faithful to reality. The onset and development of GIs and the subsequent evolution of the disk under these conditions is studied in this dissertation. In particular, the issues addressed are:

---

* The full Ph.D. dissertation is available at http://westworld.astro.indiana.edu

1. The effect of controlled cooling on the strength of gravitational instabilities; more specifically, the degree of instability, as measured by the Toomre parameter $Q$, and the propensity of the disk to fragment according to the Gammie (2001) criterion;

2. The magnitude of mass transport and the disk's growth rate, related to the angular momentum transport, as a function of cooling time;

3. Same as 1 and 2 above when the disk is cooled realistically by radiative cooling instead of a simple volumetric cooling prescription;

4. Changes in the structure of the disk when absorption of stellar radiation at its surface is added as an external heating mechanism; and

5. Changes in the properties of the optically thick and thin regions when the assumptions of the grain dust are varied; in particular, the size of the largest grains, which is related to the age of the disk if it is assumed that grains grow with time.

As much as possible, the simulations are integrated for many orbits to allow the disk to evolve well past any initial transients. In the course of its evolution, the disk adapts itself to the new cooling and heating rates and achieves a quasi-steady thermal equilibrium.

Points 1 and 2 are discussed in Chapter 3, in which three simulations, each with a different constant cooling time, are analyzed in terms of mass transport, energy evolution, and Fourier mode decomposition. Points 3 through 5 are studied in Chapter 4 with two main simulations, one heated by the star and the other not, and both subject to radiative cooling. The early stages of those two simulations are also repeated using a different maximum dust grain size.

## 2: Summary of Results

The constant cooling time simulations from Chapter 3 are referred to as $t_{cool} = 2$, $t_{cool} = 1$, and $t_{cool} = 1/4$, depending on the cooling time in outer rotation periods (ORPs, 1 ORP = 250 yr). The radiative cooling simulations in Chapter 4 are named Irrad and Shade, depending on whether stellar irradiation was included or not, respectively. All simulations include shock heating and dissipation by artificial viscosity.

The initial axisymmetric equilibrium model (section 3.1.1) is exactly the same in $t_{cool} = 2$, Shade, and Irrad. The $t_{cool} = 1$ and $t_{cool} = 1/4$ simulations started from the $t_{cool} = 2$ disk after 11.2 ORPs of evolution. Four phases are observed during the evolution of $t_{cool} = 2$. In the *axisymmetric* phase, the disk cools continuously until the internal energy reaches a minimum.

The disk becomes gravitationally unstable, develops spiral structure and expands very rapidly in the *instability* phase. The material transported outwards falls back and creates shocks that heat the disk, and so further cooling is needed to bring it back to instability. During the *adjustment* phase, the heating and cooling processes react to each other and adjust to find a balance. Eventually, the internal energy settles to a nearly constant value, cooling and heating rates become approximately the same, and the general properties of the disk stop varying significantly. At this point, the disk is in its *asymptotic* phase. The $t_{cool} = 1$ and $t_{cool} = 1/4$ simulations start after $t_{cool} = 2$ enters its asymptotic phase, so the disk goes again through a minor readjustment to the new cooling time before going back to quasi-steady equilibrium. The first three phases were also observed in Shade and Irrad. However, it is not clear that they evolved fully to the asymptotic phase, so the results mentioned in this conclusion are the final results of these simulations, but not necessarily their exact equilibrium values.

The following table shows the final values of various parameters for the five main simulations: The average Toomre $Q$ in the unstable zone, the average mass transport rate not including the instability phase, the power-law indices of the surface density and the effective temperature, the outer radius of the disk, the final value of the total internal energy, and whether the disk develops rings and fragments. For this table, the outer disk radius is taken as the radius at which the power law of the surface density breaks down. Although the final transport rate of Irrad is twice as high as Shade's, the latter disk is about 10 AU larger in radius due to the 20% higher transport rates during Shade's instability phase.

| Simulation | $Q$ | $\dot{M}$ ($M_\odot$/yr) | $\Sigma$ power | $T_{eff}$ power | $R_{disk}$ (AU) | $u$ (erg) | Rings | Fragments |
|---|---|---|---|---|---|---|---|---|
| $t_{cool} = 2$ | 1.44 | $5 \times 10^{-7}$ | 2.50 | 0.71 | 52 | $9.4 \times 10^{40}$ | Yes | No |
| $t_{cool} = 1$ | 1.43 | $10^{-6}$ | 2.47 | 0.73 | 60 | $9.9 \times 10^{40}$ | Yes | No |
| $t_{cool} = 1/4$ | 1.50 | $4 \times 10^{-6}$ | 2.48 | 0.65 | 63 | $8.4 \times 10^{40}$ | Yes | Yes |
| Shade | 1.69 | $10^{-6}$ | 2.35 | 0.87 | 54 | $1.2 \times 10^{41}$ | Yes | No |
| Irrad | 1.54 | $2.5 \times 10^{-6}$ | 2.34 | 0.69 | 44 | $1.2 \times 10^{41}$ | Maybe | No |

**Table 5-1: Summary of various parameters for the simulations in Chapters 3 and 4. The simulation $t_{cool}$=1/4 fragments under high resolution.**

## 2.1: Mass Transport and Disk Growth vs. Cooling Time

The most robust result from the constant $t_{cool}$ simulations is that the asymptotic mass transport rates observed in each case are inversely proportional to the cooling time. As a consequence of this, the disk also grows in radial extent faster with short cooling times. This is to be expected because cooling sustains gravitational instabilities, which in turn redistribute the mass and angular momentum of the disk by means of gravitational torques. The shorter the cooling time, the stronger the spiral structure (as the Fourier analysis indicates), the larger the torques, and the faster material moves within the disk.

Various spiral wave patterns ($m = 1$ to 6) appear in the unstable part of the disks, mostly between their inner and outer Linblad resonances. However, the dominant pattern is $m = 2$ and its corotation radius is the same as the radius that separates inward from outward mass transport, indicating that these modes dominate the torques that redistribute mass between about 10 and 40 AU. Thus, the mass transport is dominated by global, not local, modes and it cannot be characterized by an $\alpha$ viscosity prescription (Laughlin & Różyczka 1996; Balbus & Papaloizou 1999; cf. Gammie 2001).

The cooling times in the radiative cooling simulations are controlled by the temperature of each cell, which changes in time and space, making this relationship between mass transport and cooling times not as clear. However, typical radiative cooling times tend to be longer than a couple of ORPs, so it is expected that the asymptotic mass transport rates will be somewhat lower than those of the constant $t_{cool}$ runs. The fact that the final values for the rates in Irrad and Shade are closer to the rates of $t_{cool} = 1$ and $t_{cool} = 1/4$ is an indication that the former simulations have not reached their asymptotic phase. For all the calculations, mass transport rates are consistent with those of very young T Tau/FU Ori systems.

## 2.2: Rings, Fragmentation and Planet Formation

The formation of rings in the inner disk seems to be a common feature of these simulations. They form from mass that moves inward from the mid-disk but cannot pass through the hot, gravitationally stable innermost disk and accrete toward the central star because there is no mechanism to transport angular momentum in that zone. Therefore, the material simply accumulates and the rings grow steadily with time, faster with shorter cooling times.

Fragmentation and the production of orbiting clumps is observed when the cooling time is on the order of the local orbital time, in agreement to the result by Gammie (2001). This is only possible in the disk with the shortest cooling time, $t_{cool} = 1/4$. Due to the nature of the cylindrical grid used by the hydrodynamics code (see Pickett et al. 2003), fragmentation only happens if the grid has high azimuthal resolution. Yet, none of the handful of clumps survives more than a local orbital period, perhaps due to disruption by tidal forces, shear, or collisions with other material. If the Gammie criterion holds, the radiative cooling disks will never come close to fragmenting, even under high resolution, though this point still remains uninvestigated. One aspect to consider in particular is sharp changes in the dust opacities used in these simulations, which can facilitate fragmentation as proposed by Johnson & Gammie (2003).

Boss (2003) observes clump formation in his radiative cooling simulations. Boss' disks, which also include heating by shocks and radiative cooling just like Shade, fragment in about 360 yr. The radiative cooling times he observes are comparable to those of Shade. However, the main cooling mechanism in those disks is convection with timescales on the order of tens of years, short enough to allow for fragmentation. Convection is not detected in Shade, and certainly not expected in Irrad, whose upper layers are hotter than the midplane directly below. This discrepancy in the results may be partly due to differences in both the initial conditions and the numerical approaches. Nevertheless, this disagreement is more likely due to something more fundamental, such as differences in the treatment of boundary conditions. Fragmentation and planet formation require a very effective cooling mechanism and convection could very well be the answer if protoplanetary disks in general are proven to be convectibly unstable. No other group is presently modeling radiative cooling in 3-D, so either result awaits to be confirmed.

Though planets may not form by fragmentation of the disks from Chapters 3 and 4, planet formation by core-accretion could be facilitated by the appearance of rings. Durisen et al. (2004) explore this possibility. The idea is that the rings appear at "natural boundaries" or discontinuities in the disk structure, such as at the edge of low to high $Q$ zones, drastic changes in opacity (e.g., Johnson & Gammie 2003), state of the gas, etc. This could result in the enhancement of the local surface density by large factors, in addition to rapid drift of solids and persistent mass accumulation at the peak of the rings, both of which will potentially shorten the timescale for runaway accretion, ultimately leading to rapid planet formation. Opacity transitions at the rings due to growth of solids may also contribute to this process. The details of this hypothesis have not been fully developed, but they are being currently investigated. A hybrid scenario could be the way to approach the problem of planet formation even if GIs or core-accretion do not efficiently form planets as separate mechanisms.

## 2.3: Internal Energy and Toomre $Q$

An important result of the simulations described in this dissertation is that the disk can indeed relax to a near equilibrium state after a long evolution. Cooling and heating processes will regulate each other until the disk becomes marginally unstable, with a nearly constant $Q$. This idea has been around for several decades (Goldreich & Lynden-Bell 1965; Lin & Pringle 1987, 1990; Tomley et al. 1991, 1994; Gammie 2001), and has now been shown in global, 3-D hydrodynamics simulations. Thermal regulation is not only obvious in $Q$, but also manifests itself in an asymptotic internal energy value. For all simulations, the final value of the internal energy after the disks reached equilibrium is essentially the same, independent of cooling time. The asymptotic Toomre $Q$ is independent of cooling time for the constant cooling time simulations. The final value of $Q$ in Shade is still varying significantly in time and in radius in what is expected to be the nearly constant $Q$ zone. It is uncertain whether this quantity will approach the asymptotic values seen in the constant $t_{cool}$ simulations or remain high. Irrad has a slightly lower, nearly constant $Q$ in both time and radius.

## 2.4: Irradiation

Unfortunately, the full effect of irradiation is not included in Irrad due to limitations of the algorithm enforced by time step size requirements. The rate at which the disk heats up due to absorption of starlight has to be severely limited because the thermally unbalanced initial disk reacts too fast to the sudden energy input, which tremendously decreases the size of the computational step. Moreover, the disk is prone to bubble up and expand outside the grid in a few orbits. A quick fix is to limit the irradiation heating times to be no less than some modest fraction of an ORP, but this limiter cuts most of the heating from the star. The correct, but challenging, way to solve this problem would be to fabricate the initial model in equilibrium and slowly increase the external energy input in the 2-D hydro code, with the intent of allowing the disk to relax to the new source of heating. Nevertheless, this study shows that even a minimum amount of energy input through radiation is enough to diminish the efficiency of GIs to produce gravitational torques, likely due to changes in the boundary conditions between the irradiation and shock fronts along the instabilities at the corotation radius of the dominant spiral wave pattern.

## 2.5: Grain Sizes

The opacities used in protoplanetary disk simulations commonly assume that the dust in the disk has similar properties to interstellar medium dust, in particular, that the sizes of the grains are on the order of microns. Both simulations and observations (e.g., Widenschilling 1997; D'Alessio et al. 2001) indicate that grains tend to grow to mm and cm sizes in relatively short timescales (tens of thousands of years) in the disk environment. Varying grain sizes results in different mean opacities because the wavelength of the radiation that interacts with the grains is on the order of the size of the grains themselves. The thickness of the atmosphere changes, as well as the thickness of the radiation absorption layer, depending on the distribution of grain sizes. The simulations with larger maximum grain sizes (Chapter 4) are not carried far enough to reach even the instability phase, but it is clear already that the disks are evolving differently. For large (~ 1 mm) maximum grain size, which is more consistent with SED-fitting models (D'Alessio et al. 2001), the upper layers of the disk become more transparent to radiation from the star rather than to its own thermal radiation, causing the starlight to penetrate into the interior of the disk. This increases the net cooling times, the disk takes longer to go unstable, and it could develop a flared morphology similar to those of observed circumstellar disks (see Figure 1-1).

## 3: Relevance of this Work

Since the discovery of more than a hundred extrasolar planetary systems with characteristics quite different from the ones seen in our own Solar System, various hydrodynamic groups are racing to build planets starting from young, gravitationally unstable circumstellar disks. The most fundamental of the challenges is to prove unambiguously that gravitational instabilities can act as an efficient mechanism for gas giant planet formation. Fragmentation of the spiral GIs into protoplanetary clumps has been demonstrated by various research groups; Mayer et al. (2002) even follow a handful of these clumps for many orbits in a combination of isothermal and adiabatic calculations. However, making long-lasting (thousands of years) clumps under not-so approximate thermal conditions has not been achieved to date. It is clear that cooling and heating dominate the amplitude of the instabilities, and the most thermally realistic of the simulations in this study seem to indicate that fragmentation is not expected to happen. Of course, this is by no means the ultimate answer, but it is a definite solid step towards the inclusion of relevant thermal effects on 3-D hydrodynamic disk simulations and planet production.

As shown in SED-fitting 1+1-D models like those of D'Alessio et al. (2001), this work also concludes that the radiative cooling and irradiation heating are dependent on the dust opacities used. Actual size and spatial distributions of grains within a disk, as well as grain properties themselves, are still being heavily investigated and modeled. The entire evolution of a disk simulation that utilizes opacities in the calculation is determined by the assumptions about dust. Any disk hydrodynamics simulation is only as good as the chemistry and particulate physics it includes.

What is clear from this study is that GIs are global (disk-wide), self-regulatory disturbances and an extremely effective angular momentum and mass transport mechanism, very likely with observable consequences. For self-gravitating disks, GIs could be the main transport mechanism in the cold, optically thick mid (several to tens of AU) disk. As mentioned by Gammie (1996) and demonstrated in this study, the onset of gravitational instabilities can reproduce the mass transport rates necessary to trigger outbursts of the FU Ori type. GIs can bring large amounts of mass near the inner disk, where other transport mechanisms, such as turbulence and/or magneto-rotational instabilities, could carry this mass all the way to the star. Even while the disk is in quasi-equilibrium, GIs can expand the disk to a few hundred AU, the commonly observed size of T Tauri disks, in a time short compared with the T Tauri phase lifetime.

## 4: Final Remarks

Some of the uncertainties that remain are whether real circumstellar disks in fact become unstable, at what stage in their evolution, and how fast can this phenomenon occur. Theory certainly supports the occurrence of GIs in young, massive disks (Laughlin & Bodenheimer 1994; Yorke & Bodenheimer 1999). However, there are no direct observations yet that show us a disk succumbing to its self-gravity and giving rise to gravitational instabilities, let alone GIs fragmenting disks to form planets. Optical and near infrared imaging techniques have become good enough that structure in debris disks (age > $5\times10^6$ - $10^7$ yr) can now be resolved (Grady et al. 2001; Clampin et al. 2003). Embedded young stars and their disks will be seen in exquisite detail with the new generation of optical, near infrared and radio telescopes, such as the James Webb Space Telescope and multinational interferometer ALMA. These and other next generation telescopes will give spatial distributions of temperature, resolve shock and ionized gas regions, and other disk features such that astronomers can recognize the processes involved in disk evolution. Until then, we need to work on the theory and simulations, trying to at least agree on fundamental grounds so that we can recognize what we see when we see it. In the near future, I plan to continue my work with radiative cooling and external sources of heating in disks using the SPH code from the *N*-body Group, led by Dr. T. Quinn, at the University of Washington. One of my goals is to investigate how using a completely different approach to model similar physics can affect disk simulations and how they compare to my findings in this dissertation. Hopefully, this will contribute to the convergence of results toward a much-needed common answer.

The study of protoplanetary disks is not only an interesting problem, but also a very meaningful one to all of us as inhabitants of our Solar System. The process of formation and evolution of planetary systems in general is more complicated than we formerly imagined. Observations of extrasolar systems have taught us that characteristics we took as universal, such as low eccentricity planetary orbits or the radial sequence from terrestrial to gaseous planets, do not apply to every system. In fact, subject to observational technique biasing, the Solar System seems to be the exception rather than the rule. We have begun to understand that it is the *details* of the formation process that truly shape the way these systems evolve and even determine the habitability of future planets. I hope that my work will incite disk researchers and the astronomical community at large to investigate even further one of the most fundamental aspects of disk evolution, the thermal behavior of gas and dust, and help us unravel the intricacies of this fascinating subject of how worlds come to be and our place in this vast Universe.